OPEN ACCESS

# Labour unions under neoliberal authoritarianism in the Global South: the cases of Turkey and Egypt

Mehmet Erman Erol ᵃ and Çağatay Edgücan Şahinᵇ

ᵃDepartment of Politics and International Studies, University of Cambridge, Cambridge, UK; ᵇDepartment of Labour Economics and Industrial Relations, Ordu University, Ordu, Turkey

**ABSTRACT**

This article analyses the trajectories of organised labour in times of neoliberalism in Turkey and Egypt and their current condition under securitised neoliberal-developmentalist regimes post-2013. Neoliberal experience in these countries was marked by continuing authoritarianism, challenging the view that economic liberalisation would lead to political democratisation. One of the most important areas of neoliberal restructuring has been labour markets. In order to achieve this, struggles over organised labour were of vital importance. Dismantling the power of dissident labour unions through coercive measures and containing other sections of organised labour through authoritarian corporatist relations has been crucial in these cases.

**RÉSUMÉ**

Cet article analyse la trajectoire des syndicats dans le contexte actuel de néolibéralisme en Turquie et en Égypte, et de leurs régimes néolibéraux et développementalistes post-2013. Dans ces pays, l'expérience néolibérale est marquée par un autoritarisme persistant qui vient contredire la théorie selon laquelle la libéralisation économique mènerait à une démocratisation politique. Les marchés du travail sont l'un des points centraux de la restructuration néolibérale. Afin d'atteindre les objectifs de cette restructuration, la lutte avec les syndicats a pris une importance vitale. Le démantèlement du pouvoir des syndicats dissidents par le biais de mesures coercives, ainsi que le contrôle d'autres branches syndicales au travers de relations autoritaires et corporatistes ont joué un rôle crucial dans ces conflits.

## Introduction

Contemporary commentary on the Middle East emphasises the hostilities and competing regional orientations of Turkey and Egypt. Indeed, following the brutal 2013 coup in Egypt, diplomatic and political ties broke off as Muslim Brotherhood-backing Turkey denounced the Sisi regime as 'illegitimate'. This article, instead, focuses on the 'ties that bind'; i.e. similarities between the two regimes regarding the continuity of neoliberal authoritarianism and restructuring of labour. Approaching from a critical political economy perspective, we focus on and attempt to make sense of the condition of labour unions and the trajectories of organised struggles under neoliberal authoritarianism in these two major regional capitalist powers. The main research questions we ask are: (i) how are the historical trajectories and dynamics of neoliberal authoritarianism impacting unions in Egypt and Turkey? and (ii) what are the challenges and opportunities for the labour movement under the authoritarian and increasingly *securitised*[1] political economy structures in these two countries?

Despite significant differences, Egypt and Turkey's political-economic trajectories and the convergences thereof provide us with room for comparative investigation and analysis, from the perspective of organised labour. In Egypt, despite the repression of independent unions and the Egyption Trade Union Federation (ETUF)'s corporatist nature (Beinin 2016), labour militancy gained momentum in the 2000s as a response to Mubarak's neoliberal authoritarianism, and it led to a brief independent union movement following 2011. The discontent with neoliberalism also gave gradual rise to the political Islamist movement in Egypt, which briefly came to power before the 2013 coup. In Turkey, the military coup of 1980 aimed to 'put an end to class-based politics' (Yalman 2009) and brought about an authoritarian management of labour relations. Islamism rose against this background, both as a state strategy against class-based politics and as a reaction to inequalities created by neoliberalism. Despite the revitalisation of the labour movement in the 1990s, the 2000s saw the gradual marginalisation and co-optation of organised labour (Akçay 2021) under the political Islamist AKP-led neoliberal restructuring. Nevertheless, working-class agency played an important role in anti-neoliberal resistances such as Tekel in 2009 and Gezi in 2013. We further argue that, after the demise and defeat of popular struggles against neoliberal authoritarianism post-2013, increasingly 'securitised' regimes came to the fore in these countries, marked by repression and the containment of organised labour under *authoritarian developmentalist-cum-neoliberal* regimes of accumulation. As a result, in recent years, both countries are consistently listed in the International Trade Union Confederation's (ITUC) Global Rights Index reports (2020; 2021) as two of the 10 worst countries for workers.

This article draws upon critical engagement with the literature on neoliberalism, authoritarianism, and the political economies of Turkey and Egypt. Through the two well-defined research questions mentioned above, we sought to critically and analytically explore the trajectories and impacts of neoliberal authoritarianism and securitisation on the labour movement and unions in these two countries. We also aimed to complement this critical engagement with some interviews as supportive (secondary) material to reveal further these dynamics. For that purpose, we conducted a series of semi-structured interviews in the summer of 2021 via the snowball sampling technique. First, we contacted union organisers and officers and some academics with close ties to labour activists and unions. Through these contacts, we managed to reach some workers, an Egyptian journalist/activist and an Egyptian translator who was employed in a Turkish-owned contract manufacturing textile factory in Egypt. Due to the pandemic, interviews were conducted online. The biggest challenge for the interviews was the neoliberal authoritarian and securitised political regimes, as we were unable to get positive responses to some of our interview requests, due to the unfavourable political atmosphere against labour organisers in both countries, but especially in Egypt. This was also reflected during the interviews, as some interviewees avoided 'politically sensitive' topics.

The structure of the article is as follows. The next section develops a conceptual debate on neoliberal authoritarianism, labour and the Global South. Following this, the trajectory of neoliberal authoritarianism and labour unions in Turkey and Egypt up until 2013 are analysed. Then, a debate on labour unions under the 'securitised' neoliberal authoritarian developmentalist regimes in these countries post-2013 is pursued. An analytical conclusion follows.

## Neoliberal authoritarianism, labour and the Global South

As frequently argued, neoliberalism is a heavily contested concept. Various debates reflect differing perspectives on its definition; i.e. where it started, if it still exists or came to an end, its geographies, and the contradictions inherent in it (Hathaway 2020). Rightly defined as an 'inequality machine' (Roccu 2020, 226) and an inherently crisis-prone form of political economy which has triggered various political-economic crises leading to widespread questioning of its legitimacy (Tansel 2017), debates on neoliberalism's 'death' intensify after each crisis. This is, for example, observed in the Global South in the debates on the infamous 'Washington Consensus' after various crises in the 1990s, or the rise of 'post-neoliberalism' under 'Pink Tide' governments in Latin America as a response to the instabilities and unrest brought about by neoliberalism. The 2008 Global Financial Crisis (GFC) and the subsequent 'populist backlash' in the Global North increased these debates further (Slobodian and Plehwe 2020). Most recently, as 'the state step[ped] in to save global economies' (Wallach 2020) in the context of the serious disruption in the global economy caused by the Covid-19 pandemic, and as the idea of incompatibility of the neoliberal framework for dealing with the climate crisis grew, some have argued that 'neoliberalism is dying' (Meadway 2021). As a result, concepts such as 'state capitalism' have gained significance to define this new era of crisis and state activism (cf. Alami and Dixon 2020).

Whilst recognising that increasing and recurrent contradictions, crises and legitimacy issues have rendered 'neoliberal hegemony' less coherent, and made the state 'more visible' in the management of the economy (including our case studies Turkey and Egypt), we argue that it continues to shape states, societies and economies in new guises and 'varieties'. Approached from a labour-centred perspective, in particular, this continuity is more evident as neoliberalism essentially focuses on the 'question of labour' for competitiveness, removing barriers to capital accumulation and increased surplus value creation; hence 'restoration and strengthening class power of capital' (Harvey 2005, 54), often through authoritarian mechanisms and techniques.

The debates on the relationship between authoritarianism and neoliberalism increased especially in the aftermath of the GFC and responses of the capitalist states to that crisis (Bruff 2014; Gallo 2021). Under the contemporary form of neoliberalism, as argued by Tansel (2017, 2), oppositional social forces are disciplined, marginalised and criminalised through the capitalist states' coercive state practices; and judicial and administrative state apparatuses are also used to limit the avenues in which neoliberal policies can be challenged. These Poulantzarian/Gramscian accounts recognise that neoliberalism is intrinsically authoritarian, but these tendencies were reinforced after the GFC against the background of unpopular crisis-resolution measures.

Considering neoliberalism as an 'inherently authoritarian form of political economy' takes us to its origins in the context of the global crisis of the 1970s. In the 1970s, both in the Global North and Global South, the post-war paradigms of economic management – Keynesianism in the Global North, and developmentalism or 'peripheral Keynesianism' (Beinin 2016) in the Global South – reached their limits. The economic manifestations of the crises were generally seen in rising inflation, increased budget and balance of payment deficits, rising real wages, and slowing of the pace of global capitalist accumulation (Clarke 2005, 58). As such, the solutions were seen in privatisation, social spending cuts, 'tight' monetary policy, and reduction of barriers to capital flows and resolving the declining pace of global capitalist accumulation via internationalisation (Hanieh 2013, 14).

However, in many parts of the world, social relations were politicised and there has been a strong 'collective' opposition (i.e. by organised labour) to this restructuring. As neoliberalism and neoliberal theorists are inherently elitist and suspicious of collectivism and democracy (Gallo 2021; Harvey 2005); the neoliberal focus was on removing the democratic mass demands on the state and upholding political authority via a strong state for a free economy (Bonefeld 2017). In many parts of the world, this authoritarian 'imposition of market imperatives throughout all spheres of human activity' (Hanieh 2013, 14) was generally achieved through an assault on organised labour to curb the power of labour and restore capitalist class power.

Within this framework and with a general focus on the 'question of labour', neoliberal restructuring policies and their modalities took different forms, especially in the Global South. The role of International Financial Institutions (IFIs) and the Washington Consensus paradigm in crafting neoliberal Structural Adjustment Programmes (SAPs) implemented in debt-ridden Global South states have been emphasised by various authors. Whilst this is a salient aspect of neoliberalism in the Global South, Connell and Dados (2014, 119) outline a weakness in these 'origin stories' of neoliberalism in which 'a system of ideas generated in the global North gains political influence in the North and is then imposed on the global South'. Such an interpretation 'downplays the agency of Southern actors in the formation of the neoliberal order' (Connell and Dados 2014, 134). Seeing neoliberalism in the Global South also as a 'development strategy' (Connell and Dados 2014), 'deployed by state managers to tackle extant or budding economic and/or political crises' (Tansel 2017, 9) would prove a more nuanced understanding of neoliberalism that overcome some limitations of the perspectives originated in the global North. As such, it would also entail taking into account such features that are generally omitted in northern accounts, such as informality in the labour markets, the role of military and differing levels of welfare provision, among other things (Connell and Dados 2014).

In light of this framework, the following sections discuss the neoliberal transformation in Turkey and Egypt since the 1970s, specifically dealing with its impact on organised labour. In both cases, neoliberal authoritarianism is preferred by state managers in order to resolve the crisis of ISI/state-led political economy and overcome politicised social relations and collective demands. However, the crisis-prone and contradictory nature of neoliberalism, as well as class struggle, impeded coherent implementation of this orientation, as manifested frequently in these cases. Hence, as we demonstrate, state managers use different strategies in different periods of neoliberalism and sometimes simultaneously (from co-optation to repression and some concessions in the form of authoritarian populism) in order to control and contain the labour movement for the smooth implementation of neoliberalism as a development strategy.

## Labour unions under neoliberal authoritarianism in Turkey and Egypt

### From military dictatorship to the AKP in Turkey: neoliberalism and labour unions (1980–2002)

Turkey's transition to neoliberalism was made possible with the 1980 military coup, as a response to the crisis of ISI, political unrest and most importantly intensified class struggle due to militant labour movement. As such, the junta leaders first targeted class unionism, in line with the broader political project of neoliberalism. Turkey's most militant federation, The Confederation of Progressive Trade Unions (DİSK), and other labour organisations which opposed the state's corporatist ideology were banned, as was all strike activity. DİSK was not permitted to resume its activities until 1992. Following the authoritarian 1982 constitution, which put significant restrictions on union activity, collective labour laws were also changed in 1983. These laws (Law No. 2821 & Law No. 2822) aimed to restrict the bargaining power of the unions, both by constraining the legal steps available to them during disputes, and by widening the scope of strike bans.

The IMF-WB backed, and US-supported neoliberal authoritarian capital accumulation regime focused on export promotion, trade liberalisation and opening to the world market, which necessitated cheap, de-unionised and disciplined labour, which was secured first under the junta rule and continued in the 1980s under the conservative right-wing Motherland Party governments. As such, growth was restored but real wages in the manufacturing industry, for example, declined by 32 per cent between 1978 and 1988 (Gökten 2021, 39). However, with the increased discontent in deteriorating economic conditions and relative democratisation towards the end of the 1980s, workers in the public sector organised various protests in the spring of 1989. State workers' wave of discontent continued until the mid-1990s. The unions in the public sector mobilised their members and non-members, particularly as a reaction to the 1994 economic crisis and the following IMF programme, austerity measures and privatisation legislation. However, under the crisis conditions, these struggles could not retain the relative gains of 1989–1993 period. Instability continued to mark the 1990s with further economic and financial crises between 1998 and 2001, managed precariously under 'weak and unstable' coalition governments, anchored to IMF for credibility.

Under the auspices of the IMF, the coalition government set out to implement various neoliberal reforms from 1998 onwards, including a social security reform bill, agricultural restructuring, privatisations, international arbitration, and an overall tight disinflation programme (Marois 2012). Leaders of 15 labour organisations and confederations convened at the TÜRK-İŞ headquarters in July 1999 and officially formed the Labour Platform to fight against these reforms, insufficient increase of salaries and pensions of public servants, informal employment, and to promote democratisation and job-security. This alliance between main union confederations (TÜRK-İŞ, DİSK, HAK-İŞ, KESK, Türkiye KAMU-SEN, MEMUR-SEN)[2] was joined by the three pensioners' associations and six professional organisations. The Labour Platform organised a mass rally in Ankara in July 1999 against the IMF and government policies which was followed by a general strike with limited success in August (Koç 1999, 34–36). As it was also possible to speak of a rather statist and independent judiciary, the unions were also appealing to courts to impede large-scale privatisations in the 1990s. Hence, considering these efforts, it would not be an overstatement to argue that organised labour 'had been the most vibrant opposition force in Turkey during the 1990s' (Akçay 2021, 88).

Following the major 2001 crisis, the Labour Platform continued to oppose crisis-resolution policies and neoliberal restructuring with mass gatherings, critical reports and anti-neoliberal policy proposals. However, the success of the Labour Platform in this period was limited, and for various reasons (ideological, political, economic) its unity could not be sustained (Kutlu 2020). The 2001 crisis and the AKP's rise to power in 2002 would amount to an unprecedented neoliberal restructuring and a further decrease in the power of organised labour in the 2000s.

### *AKP, neoliberal restructuring and labour, 2002–2013*

As the AKP won a landslide victory in the 2002 elections and formed a majority government which was seen as crucial for 'neoliberal stability', the pace of neoliberal restructuring – this time under the 'post-Washington Consensus' guise and the 'democratisation' discourse – was accelerated. The AKP's credibility in economic management was secured with the EU accession process, US support, and IMF standby agreements. As such, with favourable global conditions as well, there was high economic growth, FDI and financial inflows (Marois 2012), as well as large-scale privatisations. The implications of these developments for workers, however, were not very favourable. The labour market was marked by increased flexibilisation (with the 2003 Labour Law No. 4857), stagnant wages, worsening working conditions, de-unionisation, rising unemployment due to 'jobless growth' and widespread informal employment (Erol 2019). There was also an enormous increase in deadly work accidents (or 'work murders' [Özveri 2021]) in the 'successful' industries for the new Turkey's economy, i.e. construction and shipyards.

Although there were struggles around the privatisation process in the 2000s, as anti-privatisation mass actions and lawsuits were pursued by unions in various industries, organised labour was not as effective to oppose or reverse these tendencies in this period. The attempts to hinder or cancel privatisations through legal means were not as effective as in the late 1990s either, as the courts did not want to be seen as responsible for crises like in the 1990s (Ertuğrul 2009). Similar developments were seen where job security was lost through legal regulations – such as the so-called 4/C regulation referring to an article of the Public Servants Law No. 657. With this regulation, state worker status existed only on paper and no job security was guaranteed, as manifested in the TEKEL (State-owned tobacco company) workers' case which led to their occupation of Ankara's main squares for 78 days. With a major public support for the workers, the council of state decided a suspension of execution, which served to protect workers' existing rights to an extent. This was perhaps the last major lawsuit case partly won by labour, as the AKP's (and its allies, such as the Gülenists) grip on the state and judiciary increased significantly with a major constitutional amendment package in 2010.

Debates on 'clientelism' in state-labour relations came to the fore under these circumstances and with the increasing neoliberal authoritarianism of the AKP. The clientelist relationship between unions and governments has its historical roots and goes back to the establishment of TÜRK-İŞ in 1952. Traditionally TÜRK-İŞ had an 'above party politics' which in practice generally amounted to establishing harmonious engagement with the political parties in power. Even though the AKP's preference was supporting and promoting trade union confederations with an Islamic political orientation (Özkızıltan 2019, 226–228), TÜRK-İŞ affiliated unions kept their hegemonic positions in various industries for a long time. However, AKP-supported and Islamist leaning HAK-İŞ and MEMUR-SEN's (Confederation of Public Servants Trade Unions) power and membership increased significantly towards the late 2000s. Moreover, in the 2010s, the two right-wing union confederations, TÜRK-İŞ (centre-right) and HAK-İŞ (Islamist) became increasingly convergent as both developed an increasingly pro-AKP stance. A similar argument could be made for the two right-wing public servants' confederations, Türkiye KAMU-SEN (Turkey Public Servants Unions Confederation) and MEMUR-SEN especially in the 2010s.

The rise of pro-AKP unions and the AKP's authoritarian corporatist union strategy was helpful for neoliberal restructuring; as it decreased the power of dissident labour organisations such as DİSK and KESK, and increased the membership of 'ideologically compatible' unions, especially those in the public servants' union confederations (Şahin 2021). While the total number of unionised public servants increased from 650,770 in 2002 to 1,468,021 in 2013, the total membership in the three largest public servants' confederations experienced a shift: left-wing KESK's membership decreased by 9.5 per cent from 262,348 members to 237,180 members, nationalist Türkiye KAMU-SEN's membership increased by 26 per cent from 329,065 members to 444,935 members, and Islamist MEMUR-SEN's membership increased by an astonishing 1,690 per cent, from 41,871 members in 2002 to 707,652 members in 2013 (ÇSGB 2013).

In this period, in order to further restructure unions and collective labour law in the private sector, the AKP introduced Law No. 6356 in 2012. This was presented as a 'democratisation' step which would reform the anti-democratic collective labour relations framework (Laws no 2821 & 2822), introduced by the military junta of the 1980s (Çelik 2015). With this new law, the threshold for the unions to be a party of a collective agreement and start a dispute process decreased from 10 per cent to 1 per cent, while some industries were merged and the number of industries was reduced from 28 to 20. Thus, the number of the total workers in these newly combined industries increased with the new legislation. Hence, while reducing the required threshold was presented as an important step to 'democratise' the collective labour relations, the result was to decrease the power of especially dissident unions so that they could not reach the new 1 per cent threshold, despite previously having been able to reach the 10 per cent under the prior legislation[3] (Şahin and Tepe 2018).

This phase in Turkey's political economy, then, was marked by deepening neoliberal restructuring in the labour markets with increased flexibilisation and precarity, double digit unemployment levels, stagnant wages for competitiveness, large-scale privatisations, and ongoing authoritarian restructuring of unions and collective labour relations. The AKP, however, managed to capitalise on favourable global conditions: first the booming global economy before the 2008 crash, then the QE policies of advanced capitalisms post-crash which made cheap money available for emerging markets in the Global South (Marois 2012; Gökten 2021, 46). Hence, with high growth rates, and having addressed some of the discontent of the 1990s (crises, high inflation, civil–military tensions) it managed to win subsequent elections in 2007 and 2011. It is important to note that the party also enjoyed significant working-class support through 'financial inclusion' (Akçay 2021), the neoliberal social assistance regime (Kutlu 2021) and identity politics. As such, considering these simultaneous dynamics of neoliberal restructuring and containment, some conceptualise this period as 'neoliberal populist' (Altınörs and Akçay 2022). However, this also led to new discontent and contradictions, and it alienated significant parts of the population – unemployed and precarious youth, dissident organised labour, secular urban classes, and Kurds. The manifestation of this discontent was the 2013 Gezi Uprisings. With deteriorating global conditions and declining growth following 2013 (Akçay 2021), and with the securitised neoliberal authoritarian regime following Gezi, labour politics would be further repressed.

## Labour unions and movement under Mubarak's neoliberal authoritarianism (1981–2011)

### *Neoliberal transition and Mubarak*

By the late 1960s, Egypt's state-led political economy ('Arab Socialism') became crisis-ridden, which was replaced by the policies of '*infitah*' (opening) in early 1970s under Anwar Sadat. Despite the liberalisation attempts under Sadat and an IMF Standby Agreement in 1976, neoliberal transition in Egypt took a gradual form, due to domestic and external dynamics. Domestically, labour unrest (led by rank-and-file union members) increased following the announcement of *infitah*; and the announcement of subsidy cuts following the IMF agreement led to 'bread riots'. Externally, as Beinin (2016, 25) explains, the oil boom of the 1970s was helpful, providing sufficient hard currency through increased revenues and remittances of workers working in oil-producing countries; which helped managing the balance of payment difficulties until the mid-1980s, and delaying full implementation of unpopular 'structural adjustment policies'. As such, the average growth rate in Egypt between 1975 and 1985 was 8.4 per cent. In the meantime, a more gradual restructuring was implemented, and militant left-wing activists and workers seen to be behind the discontent of the mid-1970s were repressed, while the ETUF's power was further consolidated and aligned with the regime, especially following the Trade Union Law of 1976 which prevented the emergence of independent trade unions (Hartshorn 2019, 27). Reflecting the corporatist nature of state-labour relations under Nasser, ETUF had been formed in 1957 and became increasingly tied to the state institutionally, to the extent that the state had a final say over ETUF's leadership and its president served simultaneously as Labour Minister up until the Mubarak era (Beinin 2016, 18).

Husni Mubarak became the president of Egypt After Sadat's assassination in 1981. Like Sadat, he was eager to liberalise Egypt's economy and 'intended to integrate Egypt into the world market' (Marfleet 2016, 43). However, the restructuring progressed slowly in his first decade in power, in the context of above-mentioned dynamics of easy economic growth and regime's cautious stance in light of the bread riots of the 1970s, as well as continued labour militancy of rank-and-file workers in public sector enterprises (Marfleet 2016). From 1986 onwards however, economic conditions worsened as growth significantly declined, unemployment doubled, and inflation, debt and public sector deficit increased significantly (Cammett et al. 2015, 293). In this context, Mubarak regime approached the IMF and signed a standby agreement in 1987, which collapsed the same year due to 'reluctant' implementation and efforts to reduce budget deficit. A more decisive agreement was reached with the IMF in 1991, known as Economic Reform and Structural Adjustment Plan (ERSAP).

The restructuring process which started in 1991 would shape the next two decades of Egypt's political economy. With a view to fixing imbalances and deficits, the plan demonstrated the commitment to 'an accelerated programme of privatisation, cuts in subsidies of staple foods and of fuel, and foreign trade liberalisation' (Marfleet 2016, 43), as well as labour market liberalisation/flexibilisation. An immediate focus of the programme was the enactment of the Privatization Law of 1991 (Law No 203), which initially put 314 public sector enterprises up for sale (Hanieh 2013, 50). However, there was initial reluctance toward these rapid mass privatisations; as both the ETUF and the bureaucracy were initially hesitant to endorse them, and the regime was concerned about the discontent mass layoffs could create (Hartshorn 2019, 32). Nevertheless, Egypt's privatisation performance in the 1990s stood out compared to other countries in the region and reached 4.4 billion dollars by 2000 (Hanieh 2013, 48–49). The real acceleration in privatisation and a close tie with the state and private sector, however, took place with Mubarak's appointment of Ahmed Nazif as Prime Minister in 2004 who formed a 'government of businessmen' (Ajl, Haddad, and Abul-Magd 2021, 62). Over the following five years 191 companies were sold and overall revenues from privatisation in the 2000s reached 11 billion dollars (Hanieh 2013, 50).

Another significant and strongly related area for restructuring was labour market flexibilisation or deregulation. As observed in many parts of the Global South, IFIs, governments and international capital saw so-called labour market rigidities as the main reasons for high unemployment, low productivity, and insufficient levels foreign investment, hence low growth. As such, in 1995 the Egyptian government attempted to change labour regulations so more flexibility could be achieved. Feeling the pressure of rank-and-file workers' protests (Joya 2020, 176), the ETUF tried to slow the pace of liberalisation and flexibilisation. Under pressure to accelerate privatisation and restore growth, however, more decisive attempts were made with the 2003 Unified Labour Law. This law made it easier to hire and fire workers and introduced indefinite temporary contracts (i.e. no limit on their renewals or maximum duration) (Hanieh 2013, 231; Joya 2020, 177). As such, with increased privatisation and flexibilisation, the number of low-waged, precarious and informal workers increased. It is estimated that in 2010 around 3 million Egyptian workers 'were employed under arrangements that gave employers the option to dismiss them at any moment' (Marfleet 2016, 45), and by the mid-2000s 'half of employees worked with no contract or social security coverage' (Paczynska, quoted in Joya 2020, 179). Our interviewees confirmed this point and declared that they had never worked under secure and permanent contracts, and generally worked informally. Especially 'in the private sector, a contract became some utopia, a dream' (Interview with Hossam el-Hamalawy).

### Workers and discontent

As neoliberalism deepened in the 2000s, discontent against neoliberal authoritarianism gradually increased. Despite the general liberal expectation that economic liberalisation would democratise the authoritarian state structure in the Middle East (cf. Roccu 2020; cf. Joya 2020), economic liberalisation was accompanied by stricter authoritarian rule (Marfleet 2016), and the rising discontent was increasingly repressed with state power. As Beinin (2016) notes, the protests and strikes of the mid-1980s and early 1990s ended with the violent suppression of the 1994 Misr Kafr al-Dawwar strike and paved the way for further neoliberal restructuring.

The ETUF's efforts were only limited to convincing the regime through its close ties to slow down neoliberalisation for a more gradual transformation. Despite the ETUF's initial official opposition to privatisation, after achieving some concessions from the regime in terms of rights of some workers in some privatised public sector enterprises, the confederation endorsed the privatisation process from the mid-1990s onwards (Joya 2020, 175). As pressure increased further under Ahmed Nazif's government, the ETUF leadership became even more allied to the regime. The regime's intervention into union affairs deepened to the extent that any dissident unionist was prevented from holding any influential position; with state security's involvement in the 2006 ETUF local elections preventing prominent labour activist Kamal Abu Eita's victory (Abdalla and Wolff 2020, 926). As such, there was little room for mobilisation and ETUF leadership 'opposed all but one strike during the entire Mubarak era' (Beinin 2016, 43).

From the mid-2000s, the neoliberal authoritarian orientation of the Mubarak regime, deepened neoliberalisation and worsening conditions under the Nazif government, along with the growing disconnection between ETUF leadership and rank-and-file workers and their organisations led to increased discontent and protests. There were 86 labour protests in 2003, 266 in 2004; and after the major strike (with 24,000 workers) in 2006 in Misr Spinning and Weaving Company in Mahalla, the number of strikes and protests further increased to 614 in 2007, 609 in 2008 and 700 in 2009 (Abdalla 2020, 148). The growing number of labour NGOs and independent activist organisations (such as the Center for Trade Union and Workers Services – CTUWS) also helped to draw attention to working conditions and would form the basis of an independent labour movement following the ousting of Mubarak in 2011. Unlike earlier periods, the regime was at times hesitant to confront the growing unrest with repression. In our interview, the Egyptian labour journalist and activist Hossam el-Hamalawy emphasised the impact of the strike wave since 2006 in terms of challenging the perception of an unshakeable regime. As Ajl, Haddad, and Abul-Magd (2021, 63) put it, although 'the labour movement did not directly cause the 2011 popular uprising, [it] did popularise a politics and culture of protest, delegitimising the regime.'

### Revolution, labour, brotherhood: towards the 2013 coup

The increasing social unrest in Egypt was further encouraged by the uprisings in Tunisia in December 2010. For decades, these two countries experienced similar trajectories of structural adjustment under neoliberal authoritarian regimes and were highly praised by IFIs for their 'reform' performance (Hanieh 2013). Hence, although it came as a surprise for most Western commentators and institutions, the uprisings had been in the making for years. Following the occupation of Tahrir Square on 25 January 2011 and widespread protests, Mubarak rule collapsed within weeks. For a controlled and not very radical transition, the Supreme Council of the Armed Forces (SCAF) assumed power following Mubarak's departure (Joya 2017).

Initially, for the independent labour movement, this new era opened up an opportunity 'to challenge both authoritarian state-labour relations and neoliberal economic policies' (Abdalla and Wolff 2020, 925). The ETUF's repressive position vis-a-vis the protests and ongoing strikes, and its firm support for Mubarak further deligitimised the federation. This situation gave rise to the independent labour movement that was in the making in the late 2000s, and on 30 January 2011 in Tahrir Square, the establishment of the Egyptian Federation of Independent Trade Unions (EFITU) was announced, with the participation of various organisations representing millions of workers (Beinin 2016). Kamal Abu Eita, the founder and leader of the independent Real Estate Tax Authority Union, became the leader of EFITU. After the declaration of the EFITU, in March 2011, the minister of labour for the transition government declared that the independent unions would be recognised if they register with the ministry (Abdalla and Wolff 2020, 927). Despite this, SCAF simultaneously made various efforts to limit the demands of workers and attempted to criminalise the strikes, but they continued in the following months. As Beinin wrote, 'nearly one million workers engaged in 1377 collective actions, including 280 strikes, during 2011, the largest number of actions with nearly twice as many participants as any year of the previous decade' (2016, 111).

The revolution also strengthened another actor that had been repressed during the Mubarak Era: the Muslim Brotherhood. As in Turkey's experience, during the neoliberal era Islamists had gradually increased their visibility and power especially within unorganised, rural and informal segments of society due to organised labour's and left-wing movements' failure to address their issues. As such, Islamists won the first parliamentary elections (November 2011–January 2012) and the presidential elections (June 2012), and the Brotherhood-affiliated Freedom and Justice Party's (FJP) candidate Mohamed Morsi became the president of Egypt. However, from the very beginning, Morsi's presidency reflected a continuity of neoliberal authoritarianism under an Islamist guise. In November 2012, he issued a constitutional decree granting him sweeping powers, and a presidential decree related to trade union law, allowing him to control the ETUF and also weaken the independent trade unions[4] (Abdalla and Wolff 2020). In the same month, he agreed with the IMF for a loan package of 4.8 billion dollars, which entailed austerity. Amid increasing economic woes and social and political unrest, he declared the return of a state of emergency in January 2013 (Ardovini and Mabon 2020, 469).

As Joya (2017, 209) argued, 'post-uprising Egypt was shaped by both inter-elite rivalries and by struggles from below.' Reflecting this, tensions between the military and the Brotherhood increased gradually. Simultaneously, labour unrest against Morsi rule increased, as following January 2013, there were 1972 collective actions in the first half of the year (Beinin 2016, 117). By June 2013, an anti-Islamist defacto alliance came to the fore between the military, secular forces and secular activists, as well as significant parts of the labour movement, which led to the brutal military coup that ousted Morsi and the Brotherhood from power (Hartshorn 2019; Joya 2017).

## Labour unions and movement post-2013: securitised regimes and neoliberal authoritarian developmentalism in Turkey and Egypt

### Turkey: labour unions and movement post-2013

Despite the conventional portrayal of the AKP as a democratising force in its earlier terms in power in the 2000s, legal changes to criminalise labour and social protests were already in the making with the Penal Code (2005), Criminal Procedure Code (2005), Counterterrorism Law (2006) and the Police Powers and Duties Law (2007). Many acts of labour, social and political protests, were included in an increasingly broad definition of terror crimes (Kıvılcım 2021, 201) and major strikes in various industries were banned during this period as well. These authoritarian tendencies increased further from 2013 onwards, following the Gezi Uprisings, and the power struggle between the AKP and its long allies; the Gülenists. After the June 2015 elections when the AKP lost its majority for the first time briefly (thus 'repeated' election same year), and the Gülenist coup attempt in July 2016, the party implemented more ambitious survival strategies and systematic authoritarian consolidation efforts (Akçay 2021). Forming new alliances with the nationalists such as the National Movement Party (MHP), AKP governed under a state of emergency for two years (2016–2018), a process resembling an 'exceptional state form' and culminating with the establishment of presidential system in 2018, giving sweeping powers to President Erdoğan. These developments amounted to a *defacto* 'permanent state of emergency' (Pınar 2021).

During this period, the main characteristics of the industrial relations system under the AKP's authoritarian neoliberalism; i.e. maintenance of the authoritarian nature of the collective labour law, adoption of a conservative discourse, establishment of clientelist relationships with workers and trade unions of Islamic political orientation (Özkızıltan 2019, 224–225), remained intact.[5] With further consolidation of authoritarianism and increased securitisation of politics (Pınar 2021), these tendencies were expanded and criminalisation and overt repression became widespread. As also seen in Egypt, national security and 'terrorism' was used as a justification for repressing strikes and protests and widening the scope of strike bans. Strikes and labour protests were seen as threats to the 'national' economy and unity (Pınar 2021, 37–38). In 2018, Erdoğan did not refrain from explicitly stating that the state of emergency was handy for banning strikes (Reuters 2018), and for ensuring smooth running of businesses. Reflecting the character of the *authoritarian developmentalist-cum-neoliberal* accumulation regime, the experience of workers employed in key sectors and mega construction projects such as Istanbul Airport (upon which the regime depended for sustained growth) was reflected in growing labour disputes which were drastically oppressed by security forces and law enforcement (Şahin 2021; Tuğal 2022). This criminalisation process was not only pursued for independent and militant unions but also for unions such as the more moderate Tümtis (a member of TÜRK-İŞ), which were still trying to protect workers' rights in the deteriorating economic atmosphere.

As far as the union confederations are concerned, TÜRK-İŞ continued to maintain close relations with the government in this period. In order to remain relevant and powerful, the TÜRK-İŞ leadership increasingly emphasised and relied on personal ties with government members, and especially President Erdoğan, instead of institutional channels (Interviewee 2). However, as mentioned above, some rank-and-file unions and workers were not willing to accept this stance and disturbed the corporatist framework. Hence the AKP further promoted Islamist HAK-İŞ membership which was in total harmony with the government. As a result, the member unions of TÜRK-İŞ gradually started to lose their bargaining and representative power (Şahin and Bengisu Tepe 2018, 179–180; Şahin 2021, 266–267) to HAK-İŞ, as TÜRK-İŞ's representation rate decreased to 54.68 per cent in 2021 from 71.8 per cent in 2013, while HAK-İŞ's representation rate increased to 34.37 per cent in 2021 from 16.31 per cent in 2013, and its membership increased to 711,295 in 2021 from 163,413 workers in 2013 (Interviewee 2).

This tendency of increased authoritarian corporatism and favouring pro-government union federations was also observed in the civil servants' confederations. Whilst by 2021, the total number of unionised civil servants reached 1,718,984 (1,468,021 in 2013), within this, the left-wing KESK's membership decreased to 132,225, from 237,180 in 2013 (due to purges and the securitised atmosphere post-2016), meanwhile the Islamist MEMUR-SEN's membership increased to over 1 million, from 707,652 in 2013, and the right-wing nationalist KAMU-SEN's membership remained similar (for the data see ÇSGB 2021). The absence of further increase in KAMU-SEN's membership throughout this time was perhaps attributable to its dissident position in the 2017 referendum. Thus, it is safe to argue that the alliance established between the AKP and MHP in recent years has had its own peculiar effects on public servants' confederations, as the membership proportions shift drastically.

As outlined in the introduction, the main pillars of the neoliberal authoritarian state in terms of the condition of labour put Turkey consistently among the 'Ten Worst Countries for Workers' documented by the ITUC. However, as dissatisfaction with working conditions and economic woes has grown since 2018, so have labour protests. In recent years worker protests or strikes took place in various industries and workplaces, as well as various municipalities throughout Turkey. Most recently, in the first two months of 2022, labour conflicts and workers' struggles in various industries and cities have escalated and turned into a strike wave. There were 108 strikes documented during this period. While they were generally repressed with either strike bans or police force, with the ongoing currency shocks, inflation, and rising unemployment, more labour unrest and disputes would be expected. The dissident union federation DİSK continues to work to advance workers' rights and democratisation, and various NGOs (such as ISIG Meclisi and Labour Studies Centre) continue work to draw attention to the condition of labour in Turkey. With the neoliberal authoritarian developmentalist model increasingly relying on devalued currency and cheap and disciplined labour, advancing these struggles and campaigns and organising wider segments of the working class would be needed more than ever.

### *Egypt: state and labour under Sisi*

Following the military coup in July 2013 led by the Morsi-appointed defence minister and army chief Abdel Fettah el-Sisi, another military-led transition period started. Political space started to close again and authoritarianism was consolidated with Sisi's election as president in an unfree and unfair election in 2014 (Joya 2017). As Abdelrahman notes, during his first 420 days in power, Sisi declared 236 new laws without any scrutiny, and 'many of these laws either criminalised new areas or made the penalties for already defined criminal activities more severe' (2017, 193–194). As also observed in Turkey, the definition of terrorism was expanded significantly, with a new 2015 law which defined almost anyone who acts to 'disturb public order' and 'harm national unity' as a terrorist (Hanieh 2018, 262).

When Sisi took over, the economic situation was shaky with an increased government deficit and record low Central Bank reserves (Cammett et al. 2015, 298). His regime benefitted from the international and regional dynamics that were re-shaping the Middle East. Apart from Turkey and Qatar, there was no support for the Brotherhood and the coup was welcomed by the increasingly powerful, anti-Brotherhood and Western aligned Gulf Cooperation Council (GCC) countries, especially UAE, Saudi Arabia and Kuwait. In 2013 and 2014, the amount of funds, Central Bank deposits, donations, and oil and gas subsidies from these three countries to Egypt reached over 26 billion dollars (Hanieh 2018, 260). In 2015, a conference for Egyptian development (proposed by Saudi Arabia) was held in Sharm el-Sheikh with significant participants including US Secretary of State John Kerry, IMF General Director Christine Lagarde, various prime ministers and representatives of TNCs (Roccu 2020). A further 4.5 billion dollars package provided and various projects were promised to Egypt (Hanieh 2018, 260). Finally, Egypt signed a 12 billion dollar extended fund facility agreement with the IMF in 2016, with the help of UAE (Roccu 2020, 235). Concomitantly, the Egyptian military also returned as a powerful economic actor, acting as 'entrepreneurs in uniform' (D'eramo 2021). Following 2014, with the army's lead and various public–private partnerships, 'Egypt had embarked on massive infrastructure projects such as the expansion of the Suez Canal, the building of a new capital city, the expansion of ports and highways, and mining and gas exploration' (Joya 2017, 213). As in Turkey, all of these amounted to 'authoritarian developmentalism', as the latest stage of neoliberalism (Arsel, Adaman, and Saad-Filho 2021).

These developments and orientation impacted Egyptian workers and labour movement severely, and their demands were characterised by the regime 'as attacking a struggling state, being too narrow, and threatening national unity at a time of crisis' (Stramer-Smith and Hartshorn 2021, 3). Although strikes and protests continued in the initial years of Sisi rule, their numbers gradually decreased, from 2239 in 2013 to 726 in 2016[6] (Abdalla 2020, 151). In order to repress existing strikes and independent unions further, the regime adopted a new trade union law in 2017, reinstating the ETUF's power and '*de facto* allowing the state-corporatist structure to re-establish its pre-2011 position as the main umbrella representing workers on the national level' (Abdalla and Wolff 2020, 931). Following this, all independent unions were dissolved in March 2018, and given 60 days to re-register with the ministry based on new arbitrary requirements introduced by the 2017 law (DTDA 2020). Of over 1000 independent unions, only 122 were able to re-register, and most of them remain on paper only (Interview with Hossam el-Hamalawy).

Besides these repressive obstacles, independent trade union movement's own mistakes and weaknesses played a role in their subordinate position. Following the 2013 coup, EFITU leader Kamal Abu Eita (a Nasserist) accepted the position of minister of manpower and migration under the military-led transition government. Only six months later he was ejected from the government, but his controversial statements and position during this post lost him much credibility and support among workers (Stramer-Smith and Hartshorn 2021; Marfleet 2016). Another important figure of the independent labour movement, Kamal Abbas (the founder of CTUWS), distanced himself from the regime and launched the Egyptian Democratic Labour Congress officially in 2013. EDLC had a strategy of enhancing union capacities at the local and grassroots level, but the negative consequence of this was reduced relevance at the national level. Furthermore, the fact that these organisations relied on individual labour leaders to a great extent, failed to organise increasingly precarious private sector and blue collar workers, and failed to form a unified labour movement or political alliances put limitations on their strength and success (Abdalla 2020; Abdalla and Wolff 2020). As such, with these weaknesses and with the help of the Sisi regime, the ETUF's monopoly position was re-established and it returned to its pre-uprising status, 'while the new trade union federations as well as the new trade unions have seen themselves, again, relegated to illegal organisations disregarded by the political regime' (Abdalla and Wolff 2020, 929).

## Conclusion

This article argued that Turkey and Egypt's neoliberal experience was marked by continuing authoritarianism, challenging the view that economic liberalisation would democratise the anti-democratic state structures in the Middle East. As we argued, one of the most important areas of neoliberal restructuring has been labour markets. In order to achieve this, struggles over organised labour were of vital importance. Dismantling the power of dissident unions through legal and coercive measures and containing other sections of organised labour through authoritarian corporatist and clientelistic relations has been crucial. Both Turkey and Egypt's experiences in the neoliberal era reflect these tendencies well and there is a striking continuity in these attempts.

In recent years however, these countries have witnessed unprecedented authoritarianism and have become increasingly 'securitised'. Both Turkey and Egypt's regimes have been characterised by defacto 'permanent state of emergency' and they have increasingly closed the political space. In the context of declined Western engagement with the region (Roccu 2020), and amid economic woes, both countries have turned to the Gulf in recent years, but to different sides of it. Turkey developed close political and economic ties with Qatar (as both backed Brotherhood), and Egypt became closer to the UAE and Saudi axis (anti-Brotherhood), although these solidified camps are now in the process of realignment. Complementing these, the role of the state in the economy has become more visible both in Egypt and Turkey since 2013. These developments led some accounts to argue that these regimes' political economies cannot be conceived as 'neoliberal'; rather the concept of 'state capitalism' is used to define these new characteristics of authoritarianism, securitisation and statism (Ricz 2021; Sayigh 2021; Öniş 2019).

As we have discussed in the theoretical section, neoliberalism has always been a contested concept and debates on it intensified against the background of various crises in the last couple of decades. While authoritarianism and capitalist states' role in the economy has become more visible in the context of crisis-ridden global capitalism; when approached from a labour-centred perspective there has been a striking continuity since the 1970s in terms of authoritarian management of labour power which entailed flexibility, competitiveness in the world market, and disorganisation of labour for the 'restoration and strengthening class power of capital' (Harvey 2005). For this reason, we argued that neoliberalism continues to shape states, societies and economies in new guises and varieties, and these could be observed in Egypt and Turkey's management of labour relations. As such, the political economies of these two countries post-2013 could be understood through the concept of neoliberal authoritarian developmentalism (Arsel, Adaman, and Saad-Filho 2021). More specifically, in the Turkish case, Tuğal (2022) uses the term 'neoliberal statism', and in the Egyptian case Khalil and Dill (2018) use 'statist neoliberalism' to define the recent political economy orientation of these countries. The implication of this orientation for the labour movement is a further consolidation of authoritarianism (Altınörs and Akçay 2022) and worsening conditions for labour, as documented by the ITUC. Revitalisation of organised and collective struggles is necessary in order to move toward economic and political democratisation and to challenge the misery that these regimes create. Only then can the despair turn into hope.

### *Interviews cited*

Interview 1: Hossam el-Hamalawy, Male Egyptian Labour Journalist, 44 years old. 07.09.2021.
Interview 2: Male Turkish union confederation officer, 38 years old. 17.06.2021.
Interview 3: Male Turkish HR Manager, worked in Turkish factories in Egypt for more than three years after Sisi, 30 years old. 07.07.2021.
Interview 4: Female Egyptian worker, worked at Turkish-owned and multinational textile factories in Egypt for more than six years, 28 years old, 04.07.2021.
Interview 5: Male Egyptian translator working for Turkish-owned and multinational textile factories in Egypt for more than five years, 32 years old, 07.07.2021.

### **Notes**

1. We use the concept 'securitised' to define how various socio-economic rights are increasingly included in the definition of national security (i.e. strikes and protests) to legitimise their repression, and also the more visible role of the repressive state apparatuses in the Althusserian sense (armed forces, police) in all aspects of social organisation, including the economy.
2. In terms of right to strike, the law allows state industrial (blue-collar) workers' right to strike while civil servants do not have this right; they can only conduct collective bargaining with the public employer.
3. Another similar move was the abolition of the notary-based union membership regulation, which was time and money consuming and made unionisation difficult. With the new regulation workers could use the 'e-government gateway' for membership to a union. Although seemingly a progressive step, widespread practice across Turkey has shown that it operated to transfer the membership of workers from one union to others – mostly pro-government unions – via various mechanisms including acquiring workers' e-government passwords when they start employment.
4. Although Morsi's appointment of Gibali al-Maraghi –'a younger member of the old guard' (Beinin 2013) could be conceived as a compromise, Islamist FJP affiliated figures assumed important positions within ETUF. Khaled El-Azhari became the vice-president of ETUF and was later appointed as Minister of Labour, which gave the FJP significant opportunities to control the ETUF and marginalise new independent trade unions (Abdalla and Wolff 2020, 931).

5. There are some developments during this period which are arguably not in line with this overall orientation. For example, around a million outsourced workers were re-integrated into the public sector by statutory decree No.696 in April 2018. However, such cases do not represent a contradiction with the government's union control policy. The timing of this move is noteworthy, as this reflected the AKP's populist policy ahead of the June 2018 elections. Furthermore, most of these workers were made members of unions which are either part of Hak-İş or Türk-İş confederations. It is also in line with the securitised political environment as these workers had to have 'security clearance' which were used against dissident workers.
6. It is important to note that these numbers were still higher than pre-revolution protests and strikes. They were generally repressed by the security forces. As one of our interviewees observed, when there is a labour dispute and protest within a workplace, generally military security intervened and when they resolve the dispute either by consent or force, they leave the workplace by making the workers chant 'long live Sisi' (Interviewee 3).

## Acknowledgements

We would like to thank special issue editors Bettina Engels and Alexis Roy for their guidance, comments, and patience, as well as two anonymous referees of the journal for their valuable comments. We also would like to thank Hossam el-Hamalawy and all our interviewees for their time and contributions.

## Notes on contributors

*Mehmet Erman Erol* is an Affiliated Lecturer at the University of Cambridge, Department of POLIS. His research focuses on politics of state and labour restructuring in the Middle East and North Africa.

*Çağatay Edgücan Şahin* is an Associate Professor of Labour Economics at Ordu University, Turkey. His research interests include workers self-management, solidarity economy and labour market policies in the Global South. The authors are the co-editors of *The Condition of the Working Class in Turkey: Labour under Neoliberal Authoritarianism* (Pluto Press, 2021).

## ORCID

*Mehmet Erman Erol*  http://orcid.org/0000-0001-8349-3696